\documentclass[11pt,superscriptaddress,amsmath,amssymb,nofootinbib,longbibliography]{revtex4-2}
\usepackage{graphicx}
\usepackage{dcolumn}
\usepackage{bm}
\usepackage{amssymb}
\usepackage{amsmath}
\usepackage{color}
\usepackage[normalem]{ulem}
\usepackage{dsfont}

\newcommand{\bwt}{\begin{widetext}}
\newcommand{\ewt}{\end{widetext}}
\newcommand{\newc}{\newcommand}
\newc{\hc}{\dagger}
\newc{\pd}{\partial}
\newc{\beq}{\begin{equation}}
\newc{\eeq}{\end{equation}}
\newc{\beqa}{\begin{eqnarray}}
\newc{\eeqa}{\end{eqnarray}}
\newc{\bi}{\begin{itemize}}
\newc{\ei}{\end{itemize}}
\newc{\ra}{\rightarrow}
\newc{\la}{\leftarrow}
\newc{\lra}{\longrightarrow}
\newc{\lla}{\longleftarrow}
\newc{\Lra}{\Longrightarrow}
\newc{\Lla}{\Longleftarrow}
\newc{\half}{\frac{1}{2}}
\newc{\fth}{\frac{1}{4}}
\newc{\hchecked}{\backslash\!\!\!\!\checkmark }
\newc{\del}{\delta}
\newc{\Del}{\Delta}
\newc{\gm}{\gamma}
\newc{\Gm}{\Gamma}
\newc{\lam}{\lambda}
\newc{\kap}{\kappa}
\newc{\tri}{\triangle}
\newc{\eps}{\epsilon}
\newc{\epsp}{\epsilon^\prime}
\newc{\ot}{\frac{1}{3}}
\newc{\tth}{\frac{2}{3}}
\newc{\ft}{\frac{4}{3}}
\newc{\wt}{\widetilde}
\newc{\ovl}{\overline}
\newc{\tchi}{\tilde{\chi}}
\newc{\ds}{\displaystyle}
\newc{\pmt}{\pm\!\pm}
\newc{\PL}{\hat{L}}
\newc{\PR}{\hat{R}}

\newc{\msm}{\mathrm{SM}}
\newc{\msh}{\mathrm{sh}}
\newc{\mpev}{\mathrm{PeV}}
\newc{\mtev}{\mathrm{TeV}}
\newc{\mgev}{\mathrm{GeV}}
\newc{\mmev}{\mathrm{MeV}}
\newc{\mkev}{\mathrm{keV}}
\newc{\mev}{\mathrm{eV}}
\newc{\Tr}{\mathrm{Tr}}
\newc{\nonr}{\nonumber}
\newc{\clbl}{\color{blue}}
\newc{\clg}{\color{green}}
\newc{\clr}{\color{red}}
\mathchardef\mhyphen="2D
\newc{\SL}{\not\!\!}
\usepackage{bm}
\usepackage{amsfonts}
\usepackage{xcolor}
\usepackage{amscd}
\usepackage{amsmath}

\begin{document}

\title{Exploring a Gauge Horizontal Model for Charged Fermion Masses}
\date{\today}

\author{We-Fu Chang}
\email{wfchang@phys.nthu.edu.tw}
\affiliation{Department of Physics, National Tsing Hua University, Hsinchu, Taiwan 30013, R.O.C. }

\begin{abstract}

We investigate an extension of the Standard Model (SM) incorporating a gauge \( U(1) \) horizontal symmetry that is free of anomalies. This extension introduces four additional un-Higgsed scalar doublets that do not develop vacuum expectation values, two scalar singlets, and a pair of vector-like fermionic singlets. Within this framework, the masses of third-generation charged fermions are generated through the conventional SM Yukawa interactions, while the masses of second-generation charged fermions are suppressed via a mechanism reminiscent of Froggatt-Nielsen. In contrast, the masses of first-generation charged fermions are predominantly determined by radiative corrections.

Unlike traditional implementations of the Froggatt-Nielsen mechanism, our model does not require additional colored vector or chiral fermions beyond the SM. This model provides an economical ultraviolet-complete mechanism to explain the observed patterns in charged fermion masses and Cabibbo-Kobayashi-Maskawa matrix elements. Notably, the electron electric dipole moment vanishes automatically at the two-loop level, and there is no charged lepton flavor violation to all orders. We also discuss potential experimental signatures that could distinguish this model from other \( Z' \) models, such as specific patterns in gauge boson decays and associated collider signatures.

\end{abstract}
\maketitle

\section{Introduction}
The Standard Model (SM) of particle physics is a remarkably successful framework for describing the fundamental constituents of matter and their interactions. However, persistent mysteries remain, particularly regarding hierarchical patterns in fermion masses and the associated mixing angles between the interaction and mass bases. The quark sector displays small Cabibbo-Kobayashi-Maskawa (CKM) mixing angles with a noticeable mass hierarchy \cite{PDG:2022}.

In the SM, the charged fermion mass, \( m_f \), is determined by its Yukawa coupling, \( y_f \), and the Higgs vacuum expectation value (VEV), \( v_0 = 246 \ \text{GeV} \), as \( m_f = \frac{y_f v_0}{\sqrt{2}} \). The coupling of a charged fermion to the  physical neutral component of SM Higgs boson, $h^0_{SM}$, is given by \( \frac{m_f}{v_0} \). This SM prediction for \( t, b, \tau \), and \( \mu \) has been tested with precision to a few tens of percent \cite{CMS:2022dwd,ATLAS:2022vkf}. While Yukawa coupling strengths associated with the third generation (ranging from 0.01 to unity) do not pose a puzzle in this context, an explanation is required for fermion masses linked to much smaller Yukawa couplings for the first two generations.

 Conversely, the neutrino sector features two large and one small Pontecorvo–Maki–Nakagawa–Sakata mixing angles, hierarchical mass squared differences, and minuscule neutrino masses \cite{PDG:2022}. This discrepancy in flavor structures hints at distinct mass-generation mechanisms in the charged fermion and neutrino sectors.
Additionally, it remains unclear whether neutrinos couple to $h^0_{SM}$.

Thus, as an initial step, we seek to understand the origins of charged fermion mass patterns and the intricate underlying dynamics.

Various attempts beyond the SM have sought to address flavor puzzles. Weinberg's pioneering work on radiative fermion mass generation \cite{Weinberg:1972ws} proposes that certain symmetries can nullify light fermion masses at the tree level, with masses emerging as finite higher-order effects. Subsequent investigations have explored the generation of radiative masses through loops involving new scalars and fermions, as seen in \cite{Balakrishna:1987qd, Balakrishna:1988ks, Babu:1988fn, Ma:1988qc, Babu:1990fr, Ma:1990ce}, scenarios incorporating new scalars and the top quark \cite{He:1989er, Dobrescu:2008sz}, and the introduction of new gauge bosons in extended gauge theories \cite{Weinberg:2020zba, Jana:2021tlx, Mohanta:2022seo, Mohanta:2023soi}. These approaches offer promising frameworks for understanding the hierarchies and mixing patterns observed in the SM.

Another paradigm in flavor physics is the Froggatt–Nielsen (FN) mechanism \cite{Froggatt:1978nt}, which addresses the hierarchies in Yukawa couplings by introducing an additional global \( U(1)_{FN} \) symmetry. In this framework, quarks and leptons carry different \( U(1)_{FN} \) charges, prohibiting small Yukawa couplings.
 This symmetry is spontaneously broken by the VEV of the flavon field, a scalar singlet under the SM but carrying one unit of \( U(1)_{FN} \) charge. Consequently, each small Yukawa coupling emerges from a $U(1)_{FN}$-invariant high-dimensional operator, formed by multiplying the SM Yukawa operator with \(| \Delta| \) powers of the flavon field or its complex conjugate, where \(\Delta\) is the net \( U(1)_{FN} \) charge of the SM Yukawa operator. The effective Yukawa coupling thus acquires a suppression factor of \( \left(\frac{v_{FN}}{M_{FN}}\right)^{|\Delta|} \), where \( v_{FN} \) is the VEV of the flavon field and \( M_{FN} \) is the cutoff scale for the contact interaction.

Although the FN mechanism is an elegant and simple idea, it typically necessitates the introduction of numerous heavy colored and color-neutral vector fermions, complicating the ultraviolet (UV) construction. Additionally, it suffers from the arbitrary \( U(1)_{FN} \) charge assignments. For example, since only the charge differences are relevant for achieving the phenomenologically desired suppression factors, a constant shift in every fermion's \( U(1)_{FN} \) charge would result in the same mass matrix structure.

In contrast, this paper delves into a UV-complete model that specifically avoids the introduction of any beyond Standard Model (BSM) colored vector or chiral fermions. This model seamlessly integrates radiative generation and FN mechanisms by introducing an extra gauge \( U(1)_F \) symmetry with distinct charges assigned to quarks and leptons. Choosing the abelian \( U(1)_F \) as a flavor symmetry candidate is straightforward, but implementing its gauging without introducing exotic chiral fermions presents a challenge. This difficulty stems from the stringent anomaly cancellation conditions that must be satisfied to construct realistic models.
Some discussions on finding anomaly-free \( U(1)_F \) models can be found in\cite{Allanach:2018vjg,Costa:2019zzy} (for related work, see also \cite{Dudas:1995yu, Chen:2008tc,Smolkovic:2019jow}). For additional work, see for example \cite{Kownacki:2016pmx} and references therein. Nevertheless, imposing the anomaly-free requirement serves as an invaluable constraint, effectively mitigating the arbitrary nature of \( U(1)_{FN} \) charge assignments typically observed in the original FN mechanism.

A novel, flavor non-universal, anomaly-free \( U(1)_F \) charge assignment is introduced within the proposed model, ensuring that only the third-generation quarks and lepton acquire masses at leading order. In this model, the scalar sector is extended with four heavy un-Higgsed doublets and one Higgsed singlet. The un-Higgsed doublets are responsible for the FN mechanism that generates masses for the second-generation fermions, while the Higgsed singlet facilitates the spontaneous symmetry breaking of the gauge \( U(1)_F \) symmetry. These exotic scalars also play a crucial role in the radiative mass generation of up- and down-quarks.

To generate the electron mass, we introduce a pair of vector-like singlet fermions, which are essential for the see-saw type neutrino mass generation mechanisms but are charged under \( U(1)_F \) in this context. Additionally, a charged scalar singlet is included, and together they produce the electron mass at the one-loop level.

Our numerical investigations confirm the framework's capability to replicate a realistic charged fermion mass spectrum and CKM mixings without significant fine-tuning. The subsequent sections of this paper provide detailed discussions on the explicit model (Section~\ref{sec:model}), offer example solutions that naturally reproduce the observed fermion mass spectrum and mixings while satisfying the stringent \( \Delta F = 2 \) processes constraint (Section~\ref{sec:numerical}), and consider possible experimental signatures (Section~\ref{sec:pheno}). A concise summary of the study is presented in Section~\ref{sec:summary}.

\section{Model}
\label{sec:model}
In this model, the SM fermions bear distinct $U(1)_F$ charges, with $\pm f_F, 0$ assigned to the three generations of a chiral field $f$. Where
the symbols $f=\{q, u, d, l, e\}$ represent the quark doublet, up-quark singlet, down-quark singlet, lepton doublet, and charged lepton singlet, respectively.
By adopting this charge assignment, our model conspicuously avoids the $[SU(3)^2_c] \times [ U(1)_F]$, $[SU(2)^2_L]\times [U(1)_F]$, $[U(1)^2_Y]\times [ U(1)_F]$,  $[U(1)^3_F]$, and gauge-gravity anomalies. To completely eliminate the $[U(1)_Y]\times[ U(1)^2_F]$ anomaly, it is crucial that the following condition is met:
\beq
q_F^2+d_F^2+e_F^2=2 u_F^2+l_F^2\,.
\eeq

After meticulously examining all potential integer values for $f_F$ up to $9$ and considering the charged fermion mass pattern, we have refined the plausible solutions to a select few. One of these realistic solutions is presented in Table~\ref{table:SMparticle}.
\begin{table}[htb]
\begin{center}
\begin{tabular}{|c||ccccccccc|cccccc|c|}\hline
     &   \multicolumn{9}{c|}{SM quark }   &   \multicolumn{6}{c|}{SM lepton }  &  \multicolumn{1}{c|}{SM Higgs}\\ \hline
 Symmetry $\backslash$ Fields& $Q_1$ & $Q_2$ & $Q_3$ & $u_1$ & $u_2$ & $u_3$
  & $d_1$ & $d_2$ & $d_3$
  & $L_1$ & $L_2$ & $L_3$ & $e_R$ & $\mu_R$ & $\tau_R$ & $H$   \\\hline\hline
$U(1)_F$ & $2$  & $-2$ & $0$ &$1$ & $-1$ & $0$ & $3$  & $-3$   & $0$   & $-6$  & $6$ & $0$ &$-5$ & $5$ & $0$ & $0$ \\
\hline
\end{tabular}
\caption{The SM fields and their quantum numbers under the gauged $U(1)_F$ symmetry. }
\label{table:SMparticle}
\end{center}
\end{table}

To streamline the generation of charged fermion masses, the scalar sector of this model is extended to include four un-Higgsed doublets and one Higgsed singlet, as shown in Table~\ref{table:newparticle}. These elements play a pivotal role in the FN mechanism responsible for generating the masses of the second-generation fermions, with the masses of the un-Higgsed doublets serving as the cutoff scale in FN.
The scalar field \( S_1 \) is crucial for the SSB of \( U(1)_F \). Once \( S_1 \) acquires a VEV, denoted as \( \langle S_1 \rangle = v_1 / \sqrt{2} \), the associated new gauge boson gains mass, which is given by \( M_F = g_F v_1 \). As will be demonstrated, the exotic scalars play a significant role and are integrated into the radiative mass generation process for up- and down-quarks.
For the generation of the electron mass, we introduce a pair of vector-like singlet fermions, $N$, and a charged scalar singlet, $C_4$, aiming to generate the electron mass at the one-loop level. The vector fermionic singlet is characterized by a tree-level Dirac mass term \( M_N \bar{N} N \).

\begin{table}[htb]
\begin{center}
\begin{tabular}{|c||c|cccccc|}\hline
       &   \multicolumn{1}{c|}{ New Fermion}  &  \multicolumn{6}{c|}{ New Scalar}\\\hline
 Symmetry$\backslash$ Fields & $N_{L,R}$
  & $H_1 $ & $H_+$ & $H_{-} $ & $H_{3} $& $S_1$  & $C_4^+$  \\\hline \hline
$SU(2)_L$ &  $1$& $2$ & $2$  & $2$  & $2$ & $1$& $1$\\
$U(1)_Y$  & $0$ &  $\half$  & $\half$ & $\half$   & $\half$  & $0$& $1$\\
$U(1)_F$ & $-9$ &  $1$ & $2$ &$-2$ & $-3$ & $1$ & $-4$\\
\hline
\end{tabular}
\caption{New field content and quantum number assignments under the SM gauge symmetries $SU(2)_L \otimes U(1)_Y$, and the gauged lepton numbers $U(1)_F$. Among the exotic scalars, only $S_1$ develops a nonzero vacuum expectation value.}

\label{table:newparticle}
\end{center}
\end{table}

It is straightforward to write down the full Lagrangian for the scalar sector, and we have no new insights to offer regarding the physics of heavy scalars. Instead, we will focus exclusively on the relevant terms that address the flavor puzzle.

The complete gauge-invariant Yukawa interaction between pairs of SM fermions can be expressed as
\beqa
\label{eq:all_Yukawas}
&-& \overline{(Q_1, Q_2, Q_3)} \left(
                                               \begin{array}{ccc}
                                                 0& y^u_3 \wt{H_{3}}&y^u_-\wt{H_{-}}\\
                                                 0& y^u_1 \wt{H_{1}}& y^u_+ \wt{H_{+}}\\
                                                 y'^u_1\wt{H_{1}}&0& y_t \wt{H}\\
                                               \end{array}
                                             \right)
                                             \left(
                                               \begin{array}{c}
                                                 u_1 \\u_2\\u_3
                                               \end{array}
                                             \right)
                -   \overline{(Q_1, Q_2, Q_3)} \left(
                                               \begin{array}{ccc}
                                                 0& 0&y^d_+ H_{+}\\
                                                 0&y^d_1 H_{1}& y^d_- H_{-}\\
                                                 y^d_3 H_{3}&0& y_b H\\
                                               \end{array}
                                             \right)
                                             \left(
                                               \begin{array}{c}
                                                 d_1 \\d_2\\d_3
                                               \end{array}
                                             \right)\nonr\\
 &-&    \overline{(L_1, L_2, L_3)} \left(
                                               \begin{array}{ccc}
                                                 0& 0& 0\\
                                                 0& y^e_1 H_{1}&0\\
                                                 0&0& y_\tau H\\
                                               \end{array}
                                             \right)
                                             \left(
                                               \begin{array}{c}
                                                 e_R \\ \mu_R\\ \tau_R
                                               \end{array}
                                             \right)    + H.c.
\eeqa

\begin{figure}
\centering
\includegraphics[width=0.7\textwidth]{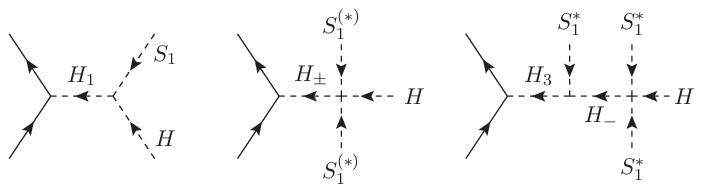}
\caption{Feynman diagrams illustrating the scalar implementation of the Froggatt–Nielsen mechanism.}
\label{fig:S_FN}
\end{figure}

The origin of third-generation masses is intrinsically linked to the Yukawa interactions present in the SM. Given the mass hierarchy observed among the top quark, bottom quark, and tau lepton, a two-order-of-magnitude distinction in Yukawa couplings within the same category is considered `natural'.

Additionally, exotic un-Higgsed doublets serve a crucial function as mediators in FN mechanisms, as illustrated in Fig.\ref{fig:S_FN}.
At lower energies, much less than $ M_{1,3,\pm}$, these interactions lead to effectively suppressed Yukawa couplings to the SM Higgs, expressed as:
\beq
y_1^{eff} \simeq y_1 \frac{v_1 \mu_{0,1,s}}{M_1^2}\,,\;
y_{\pm}^{eff} \simeq y_{\pm}\, \lambda_{\pm,0,s,s}\, \frac{v_1^2}{M_\pm^2}\,,\;
y_3^{eff} \simeq y_3\, \lambda_{-,0,s,s}\, \frac{v_1^3 \mu_{-,3,s}}{M_-^2 M_3^2}\,.
\label{eq:FNSF}
\eeq
The relevant vertices for Yukawa couplings, dimensionful scalar cubic couplings, and dimensionless scalar quartic couplings in the expression are clearly identified from the accompanying diagrams and their subscripts. Specifically, \( H_1 \) is responsible for generating the diagonal mass matrix elements for the second-generation fermions. The roles of \( H_{\pm} \) are particularly noteworthy, as these fields collectively provide the necessary off-diagonal mass matrix elements to accommodate the observed CKM  mixings.

In the context of the FN mechanism, the charged fermion mass matrices are given by:
\begin{equation}
M_u \simeq \left(
\begin{array}{ccc}
0 & \star^u_3 & \star^u_- \\
0 & \star^u_1 & \star^u_+ \\
\star^u_{1'} & 0 & m_t \\
\end{array}
\right)\,,\quad
M_d \simeq \left(
\begin{array}{ccc}
0 & 0 & \star^d_+ \\
0 & \star^d_1 & \star^d_- \\
\star^d_3 & 0 & m_b \\
\end{array}
\right)\,,\quad
M_e \simeq \left(
\begin{array}{ccc}
0 & 0 & 0 \\
0 & \star^\ell_1 & 0 \\
0 & 0 & m_\tau \\
\end{array}
\right)\,,
\end{equation}
where each \(\star\) denotes a suppressed mass matrix element of the form \(\star = y^{\text{eff}} v_0 / \sqrt{2}\). The effective Yukawa coupling \(y^{\text{eff}}\) arises from the FN mechanism, with the subscripts indicating the associated FN mediator.
Notably, the up-type and down-type quark mass matrices contain three and four structural zeros\cite{Fritzsch:1977vd,Fritzsch:1979zq,Xing:2020ijf}, respectively . At tree level, these mass matrix structures yield see-saw suppressed masses for the up and down quarks as:
\beq
|m^{tree}_u| =  \frac{ | \star^u_{1'} (\star^u_{1} \star^u_- -\star^u_{3} \star^u_+)| }{m_c m_t}\,, \quad | m^{tree}_d | =  \frac{ |\star^d_3 \star^d_1 \star^d_+|}{m_s m_b} \,,
\eeq
while \(m_e = 0\).

However, when one-loop corrections are considered, all seven structural zeros in the quark mass matrices are lifted, with each element receiving a nonzero but small contribution from quantum corrections. Although the see-saw-like lightest quark mass eigenvalue is nonzero at tree level, the quantum contributions to the diagonal \((1,1)\) component, denoted as \(\delta^{u,d}_{11}\), become more significant in determining \(m_u\) and \(m_d\) once the structural zeros at the \((1,1)\) positions are lifted. They become:
\beq
|m_u| \simeq \left| \delta^{u}_{11} - \frac{\star^u_{1'} \star^u_-}{m_t} \right|\,, \quad |m_d| \simeq \left|\delta^d_{11} - \frac{\star^d_3 \star^d_+}{m_b}\right|\,,
\eeq
where we  assume that \(|\star^u_1| \sim m_c\), $|\star^d_1| \sim m_s$, \(|\star^u_3| \ll m_c\), and ignore higher-order terms, \({\cal O}(\delta^2)\), from quantum corrections to other mass matrix elements. Note that \(H_3\) is crucial to ensure \(m_d \neq 0\) at both tree and loop levels (which will be discussed further below).

\begin{figure}
\centering
\includegraphics[width=0.7\textwidth]{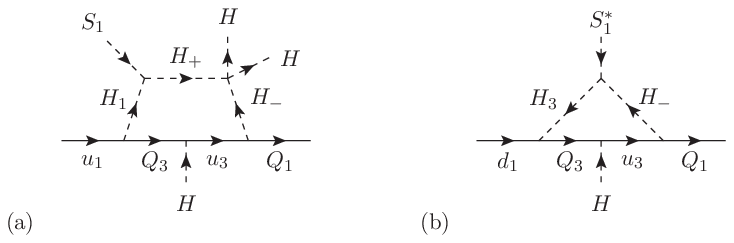}
\caption{Feynman diagrams (in the interaction basis) for the dominant 1-loop contributions to the up-quark and down-quark masses.}
\label{fig:u_d_mass}
\end{figure}

The one-loop diagrams depicted in Fig.~\ref{fig:u_d_mass} provide dominant finite contributions to the \((1,1)\) entries of \(M_u\) and \(M_d\).
Both up- and down-quark masses involve the top quark in the loop. The up-quark mass is generated using the neutral components of the exotic Higgs doublets, while the charged components contribute to the down-quark mass. Conversely, the structural zeros in the charged lepton mass matrix persist due to the accidental lepton flavor numbers, which will be elaborated upon in subsection~\ref{sec:leptonNum}.

Upon closer inspection of the diagrams in Fig.~\ref{fig:u_d_mass}, it becomes evident that the 1-loop quantum corrections to the up and down-quark masses are given by
\begin{eqnarray}
\delta m_u^{1-loop} &\simeq& m_t \frac{y^u_- y'^u_1}{16\pi^2} \frac{\mu_{+,1,s} \lambda_{+,-,0,0}}{\Lambda^4} v_1 v_0^2 \ln \frac{\Lambda^2}{v_0^2}\,, \\
\delta m_d^{1-loop} &\simeq& -m_t \frac{y^d_3 y^u_-}{16\pi^2} \frac{\mu_{-,3,s}}{\Lambda^2} v_1 \ln \frac{\Lambda^2}{v_0^2}\,,
\label{eq:m_del_d_1loop}
\end{eqnarray}
where $\Lambda$ denotes the cut-off scale, and the other symbols in the expressions are self-explanatory. Note that the minus sign in the down-quark mass arises from the tilded conjugate $\tilde{H}_-$.

Interestingly and unexpectedly, the model automatically yields \(\delta m_u^{1\text{-loop}} < \delta m_d^{1\text{-loop}}\), which is contrary to the observed pattern where \(m_t > m_b\) and \(m_c > m_s\).
However, $\delta m_u^{1-loop}$ experiences significant suppression by a factor of \( m_t v_0^2 / v_1^3 \), rendering it four orders of magnitude too small.  Conversely, if one splices the SM Higgs coupled to top quark with the one from the quartic coupling vertex, the resulting 2-loop contribution undergoes much less suppression and yields a value in the right ballpark:
\begin{equation}
\delta m_u^{2-loop} \simeq  m_t\, \frac{ y^u_- y'^u_1 \lambda_{+,-,0,0}}{(16\pi^2)^2} \frac{\mu_{+,1,s} v_1}{\Lambda^2} \left(\ln \frac{\Lambda^2}{v_0^2}\right)^2\,.
\label{eq:m_del_u_2loop}
\end{equation}

Now, let us consider the electron mass generation. The interactions involving \( N \) and \( C_4 \) are entirely determined by their quantum numbers. The most general gauge-invariant Yukawa interaction associated solely with first-generation leptons, is given by:
\begin{equation}
\label{eq:Lag_e}
\mathcal{L} \supset - y^e_{3} \overline{N_R} H_{3}^T (i\sigma_2) L_1 - y^e_c \overline{N}_L e_{R} C_4^+ - M_N \overline{N_R} N_L + \text{h.c.}
\end{equation}

\begin{figure}
\centering
\includegraphics[width=0.27\textwidth]{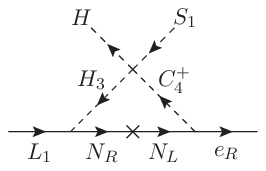}
\caption{Feynman diagrams (in the interaction basis) for the radiative mass generation of the electron.}
\label{fig:e_mass}
\end{figure}

Additionally, there is a relevant symmetry-allowed scalar interaction given by \( \lambda_e H_3^T (i\sigma_2) H C_4^- S_1^* \), where \( \lambda_e \) is the quartic coupling. The 1-loop diagram, as shown in Fig.~\ref{fig:e_mass}, results in the following correction to the electron mass:
\begin{equation}
\delta m_e \simeq M_N \frac{y^e_3 (y^e_C)^*}{16\pi^2} \frac{\lambda_e v_1 v_0}{\Lambda^2} \ln \frac{\Lambda^2}{v_0^2}\,.
\end{equation}

The connection between the parameters in the above equation and the corresponding vertices in the diagram is readily apparent.
A more detailed consideration of $\delta m_e $ will be given in Sec.~\ref{sec:delta_m_E}.

\subsection{Lepton Flavor Symmetries and Charged Lepton Flavor Violation}
\label{sec:leptonNum}

Upon careful examination of the lepton-related interaction terms in the Lagrangian, as detailed in Eqs.~(\ref{eq:all_Yukawas}) and~(\ref{eq:Lag_e}), it becomes evident that the model exhibits three accidental global \( U(1) \) lepton flavor symmetries. Specifically, \( L_1 \), \( e_R \), and \( N_{L,R} \) each carry one unit of electron number, while \( \{ L_2, \mu_R \} \) and \( \{ L_3, \tau_R \} \) carry one unit of muon and tau number, respectively. All fields other than leptons and \( N_{L,R} \) are neutral under these lepton numbers. These global \( U(1) \) lepton symmetries are preserved even after electroweak symmetry breaking, leading to  a see-saw plus FN suppressed electron neutrino mass, $m_{\nu_e}\sim (y^e_{3,eff} v_0)^2/M_N$\footnote{We estimate a sub-eV range for \(m_{\nu_e}\) by applying the FN suppression factor for the \(H_3\) Yukawa, as discussed in the next section. Assuming \(M_N \sim \text{PeV}\) and \(y_e^3 \sim 0.1\), we obtain this ballpark value.}
, massless $\nu_{\mu,\tau}$, and ensuring compliance with stringent limits on charged lepton flavor violation.

In contrast to the quark sector, the \( U(1) \) lepton flavor symmetry in this model preserves the structural zeros in the charged lepton mass matrix,  preventing them from being lifted by quantum corrections.

\section{Numerical Analysis and Implications}
\label{sec:numerical}
In this section, we begin by presenting a simplified benchmark solution that provides a realistic quark mass spectrum and CKM mixings. We then proceed to a numerical exploration of the broader model parameter space.

Given that the model necessitates a mass scale significantly higher than the electroweak scale, it is crucial to account for the renormalization group evolution (RGE) of the quark masses and CKM mixings. The parameters selected for fitting are evaluated at the energy scale \(\mu =1\,\mpev= 10^6\,\mgev\) \cite{Huang:2020hdv}, as detailed below:
\beqa
m_t=114.94(87)\mgev\,,\; m_c=0.399(12)\mgev\,,\; m_u=0.79(14)\mmev\,,  \nonr\\
m_b=1.734(20)\mgev\,,\; m_s=0.0345(30)\mgev\,,\;m_d=1.73(12)\mmev\,,\nonr\\
m_\tau=1.7873(21)\mgev\,,\; m_\mu=0.10520(13)\mgev\,,\; m_e=0.49936(75)\mmev
\label{eq:fittingQMass}
\eeqa
The RGE running also impacts $V_{us}, V_{cd}, V_{tb}, V_{cs}$, and $V_{ud}$ \cite{Barger:1992pk}. At $\mu=1\ \mpev$, the strengths of $V_{cb}, V_{ub}, V_{ts}, V_{td}$ experience a roughly $7.6\%$ enhancement compared to their values at low energy, while the enhancement for the Jarlskog invariant\cite{Jarlskog:1985ht} is approximately $15.9\%$ \cite{JuarezWysozka:2002kx}. Utilizing the CKM matrix elements at low energy\cite{PDG:2022} and the aforementioned enhancement factors, we adopt the following values for the fitting at $\mu=1\, \mpev$:
\beq
 |V_{us}|=0.2250(7)\,,\; |V_{cb}|=0.0450(9)\,,\; |V_{ub}|=0.00397(12)\,,\; J=3.57(17)\times 10^{-5}\,.
 \label{eq:fittingCKM}
\eeq

\subsection{A Simplified Benchmark Solution}

To simplify the discussion and reduce the number of free parameters, we consider a scenario where only \( y^u_+ \) is complex. This assumption effectively eliminates CP violation in kaon mixing and addresses the CKM weak CP phase.
For illustration, consider the following benchmark point:
\begin{eqnarray}
M_u &\simeq& \left(
\begin{array}{ccc}
4.71 \times 10^{-4} & 0.0888 & 0.0407 \\
2.80 \times 10^{-4} & 0.392 & 0.331 \, e^{0.437 i} \\
1.078 & 2.25 \times 10^{-3} & 114.67 \\
\end{array}
\right) \, \text{GeV}\,, \nonumber \\
M_d &\simeq& \left(
\begin{array}{ccc}
1.59 \times 10^{-3} & 1.78 \times 10^{-4} & 0.016 \\
1.05 \times 10^{-4} & 0.0361 & 0.0796 \\
-4.549 \times 10^{-3} & 9.59 \times 10^{-4} & 1.722 \\
\end{array}
\right) \, \text{GeV}\,.
\end{eqnarray}
These quark mass matrices can be generated by the following set of model parameters:
\beqa
v_1 &=& 2 \,\mpev\,, \quad M_+ = 26.45 \,\mpev\,, \quad M_- = 22.75 \,\mpev\,, \quad M_3 = 12.48 \,\mpev\,, \quad M_1 = 22.24 \,\mpev\,, \nonr\\
\lambda_{-,0,s,s} &=& 0.527\,, \quad \lambda_{+,0,s,s} = 0.696\,, \nonr\\
\mu_{0,1,s} &=& 2.97 \,\mpev\,, \quad \mu_{-,3,s} = 11.22 \,\mpev\,,  \quad (\lambda_{+,-,0,0} \times \mu_{+,1,s}) \simeq 4.56 \,\mpev\,, \nonr\\
y^u_3 &=& 0.870\,, \quad y^u_+ = 0.477 e^{0.437 i}\,, \quad y^u_- = 0.057\,, \quad y^u_1 = 0.187\,, \quad y'^u_1 = 0.515\,, \nonr\\
y^d_3 &=& -0.0445\,, \quad y^d_+ = 0.0233\,, \quad y^d_- = 0.112\,, \quad y^d_1 = 0.0172\,,\quad y^e_1 = 0.0502\,.
\label{eq:BENCHM_Y}
\eeqa
Observe that  $\delta m_u$ cannot constrain the two individual parameters but only their product, $\lambda_{+,-,0,0} \times \mu_{+,1,s}$.
Note that the absolute values of the Yukawa coupling strengths to the exotic Higgs doublets are approximately in the ranges of \([0.1, 1.0]\) for the up-quark sector and \([0.01, 0.2]\) for the down-quark and charged lepton sectors. This substantial reduction significantly narrows the five-order-of-magnitude differences observed in the SM Yukawas.
Notably, the muon Yukawa coupling to \( H_1 \) exceeds the SM tau Yukawa coupling, with \( y^e_1 = 0.0502 \) significantly larger than \( y^\tau_{\text{SM}} = 0.0072 \). Similarly, the Yukawa coupling for the strange quark, \( y^d_1 \), surpasses the SM bottom quark Yukawa coupling, \( y^b_{\text{SM}} \).

Regarding quantum corrections, we focus on the \((1,1)\) components of the two mass matrices. Equations (\ref{eq:m_del_u_2loop}) and (\ref{eq:m_del_d_1loop}) are used for estimation, as they provide the dominant contributions to the up and down quark masses. Quantum corrections to other tree-level structure zeros are assumed to be controlled by additional model parameters and are not expected to have significant effects.

This approach results in a fit that aligns well with the previously mentioned values listed in Eq.~(\ref{eq:fittingQMass}) and Eq.~(\ref{eq:fittingCKM}). It is noteworthy that the mass matrix elements associated with the FN mechanism require only an order-one tuning in their Yukawa couplings. Additionally, the elements that arise through radiative processes are generally small and exhibit a hierarchy of less than order one among themselves.

Using the benchmark values mentioned above, we have determined the scalar FN suppression factors for the new Higgs doublets relative to the SM Yukawa couplings. The hierarchical suppression factors,  see Eq.(\ref{eq:FNSF}), are \(\frac{v_1 \mu_{0,1,s}}{M_1^2} = 0.012\) for \(H_1\), \(\lambda_{\pm,0,s,s} \times \frac{v_1^2}{M_\pm^2} \simeq 0.004\) for \(H_{\pm}\), and \(\lambda_{-,0,s,s} \times \frac{v_1^3 \mu_{-,3,s}}{M_-^2 M_3^2} = 0.00059\) for \(H_3\), respectively.

\subsection{Constraints from Flavor-Changing Neutral Currents}
One must account for the stringent constraints imposed by Flavor-Changing Neutral Currents (FCNCs) mediated by the new gauge boson. In the mass basis, the couplings to the new gauge boson are given by:
\begin{equation}
Q^q_{L/R} = (V^q_{L/R})^\dag \cdot \text{diag}\{ q^q_{L/R}, -q^q_{L/R}, 0 \} \cdot V^q_{L/R}\,,
\end{equation}
where \( V^q_{L/R} \) are the mixing matrices that diagonalize the quark mass matrix.

Using the benchmark solution, the gauge couplings in the quark mass basis at \(\mu = M_F\) are as follows:
\begin{eqnarray}
Q^d_L &\simeq& \left(
\begin{array}{ccc}
1.999 & -0.0195 & 0.0192 \\
-0.0195 & -1.999 & 0.0921 \\
0.0192 & 0.0921 & -0.00409 \\
\end{array}
\right)\,, \\
Q^d_R &\simeq& \left(
\begin{array}{ccc}
2.999 & 0.0537 & -0.00783 \\
0.0537 & -2.999 & -0.0046 \\
-0.00783 & -0.0046 & 1.37 \times 10^{-5} \\
\end{array}
\right)\,, \\
Q^u_L &\simeq& \left(
\begin{array}{ccc}
1.805 & 0.861 \, e^{-0.44 i} & 0.0019 \, e^{2.56 i} \\
0.862 \, e^{0.44 i} & -1.805 & 0.0055 \, e^{0.023 i} \\
0.0019 \, e^{-2.56 i} & 0.0055 \, e^{-0.023 i} & -1.63 \times 10^{-5} \\
\end{array}
\right)\,, \\
Q^u_R &\simeq& \left(
\begin{array}{ccc}
0.999 & 0.0137 \, e^{0.253 i} & -0.0094 \\
0.0137 \, e^{-0.253 i} & -0.999 & 9.3 \times 10^{-5} \, e^{2.78 i} \\
-0.0094 & 9.3 \times 10^{-5} \, e^{-2.78 i} & 8.84 \times 10^{-5} \\
\end{array}
\right)\,.
\end{eqnarray}

The most stringent experimental constraints arise from $K^0-\bar{K^0}$, $B^0-\bar{B^0}$, and $D^0-\bar{D^0}$ mixings. Consequently, our primary focus is on the relevant 4-fermi operators.
In conventional notation, as detailed in \cite{Ciuchini:1998ix}, the only non-zero Wilson coefficients at $M_F$ are associated with the flavor-crossing gauge couplings are:
\beq
C_{ij}^1 =  {[(Q_L)_{ij}]^2 \over v_1^2}\,,
\wt{C}_{ij}^1 =  {[(Q_R)_{ij}]^2 \over v_1^2}\,,
C_{ij}^5= -4{(Q_L)_{ij} (Q_R)_{ij} \over v_1^2}\,,
\eeq
These coefficients are independent of the coupling constant since $M_F= g_F v_1$, and only the VEV $v_1$ is relevant.

However, at the corresponding meson scale, non-vanishing $C_{ij}^4$ emerges from the RGE running down from $M_F$ \cite{Ciuchini:1998ix,Smolkovic:2019jow}.
In this benchmark, the Wilson coefficients for $K^0\mhyphen \overline{K^0}$ mixing at $\mu=M_F$ are given by:
\beq
C^1_K(M_F) = 3.81\times 10^{-4} v_1^{-2}\,,\; \wt{C}^1_K(M_F) =  2.89\times 10^{-3} v_1^{-2}\,,\; C^5_K(M_F) = 4.19\times 10^{-3} v_1^{-2}\,.
\eeq
The RGE running yields:
\beqa
C^1_K(\mu_K) = 2.51\times 10^{-4} v_1^{-2}\,,\; \wt{C}^1_K(\mu_K) =  1.89\times 10^{-3}  v_1^{-2}\,,\nonr\\
 C^5_K(\mu_K) =3.65\times 10^{-3} v_1^{-2}\,,\;  C^4_K(\mu_K) =  1.18\times 10^{-2} v_1^{-2}\,,
\eeqa
where $\mu_K =2 \, \text{GeV}$.

We have established that the most stringent constraint is \( |\Re C^4_K| < 3.6 \times 10^{-15} \ \text{GeV}^{-2} \) \cite{UTfit:2007eik}, corresponding to \( v_1 > 1.81 \ \text{PeV} \). Given the sizable \((12)\) component of \( |Q^u_L| \) in this benchmark, we carefully verified that, even after considering the RG running \cite{Becirevic:2001jj,UTfit:2007eik} to \( \mu_D = 2.8 \ \text{GeV} \), the constraint \( |C^4_D(\mu_D)| = 0.109 v_1^{-2} \ \text{GeV}^{-2} < 4.8 \times 10^{-14} \ \text{GeV}^{-2} \) \cite{UTfit:2007eik} still imposes a slightly weaker bound of \( v_1 > 1.51 \ \text{PeV} \) compared to \( \Re C^4_K \). Therefore, it is permissible to select \( M_F = 1.0 \ \text{PeV} \) with \( v_1 = 2 \ \text{PeV} \) and \( g_F = 0.5 \).

Nevertheless, the new gauge coupling strength is an unknown parameter. If one assumes a smaller \( g_F \), the mass of the new gauge boson is lighter. For example, \( M_F \simeq 3 \ \text{TeV} \) if \( g_F \simeq 0.0015 \), and it may be probed in a future high-energy \( e^- e^+ \) collider \cite{Chang:2018nid, Chang:2022eft, Chang:2022pue}. For more discussion, see Section~\ref{sec:ZX_pheno}.

\subsection{Exploring Complex Yukawa Interactions}
To gain deeper insights into the behavior of the model, we extend our previous analysis by conducting a thorough numerical scan of the model parameters. This comprehensive study enables us to assess how naturally the model accommodates realistic data.

We adopt the benchmark values for \( v_1 \), \( M_\pm \), \( M_3 \), and \( M_1 \), along with the corresponding scalar FN suppression factors derived from Eq.~(\ref{eq:BENCHM_Y}). For the \((33)\) components, we keep \( y_t \) and \( y_b \) within \(10\%\) of their SM values. For other Yukawa couplings, we allow the absolute values to vary between \(0.1\) (\(0.01\)) and \(1.0\) (\(0.1\)) for the up (down) quark sector, each with an arbitrary complex phase. Each tree-level structural zero element is constrained to have an absolute value of less than \(2 \ \mmev\) with a random phase.

\begin{figure}[htb]
\centering
\includegraphics[width=0.4\textwidth]{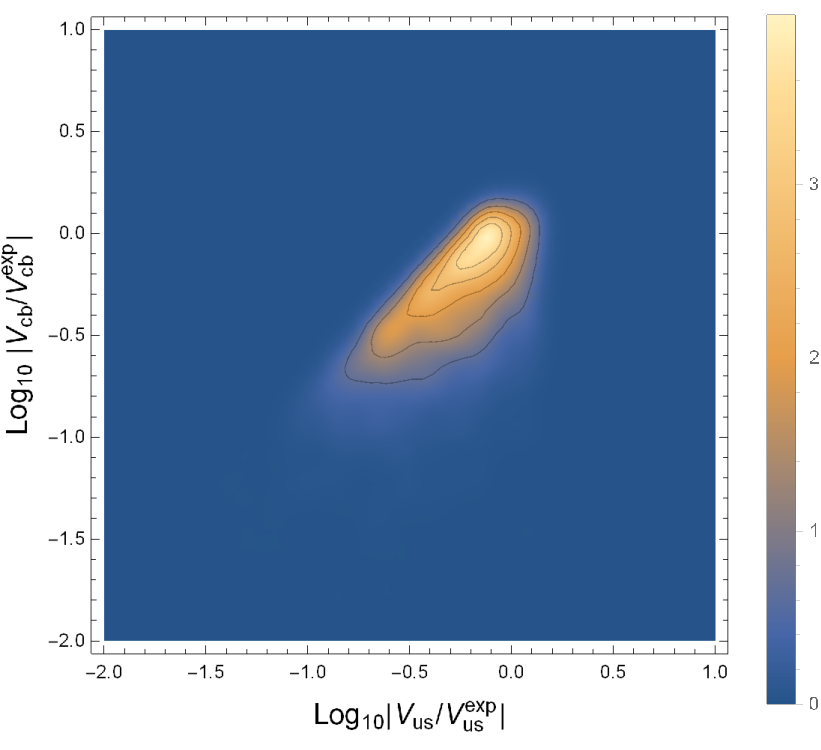}
\hspace{0.5cm}
\includegraphics[width=0.4\textwidth]{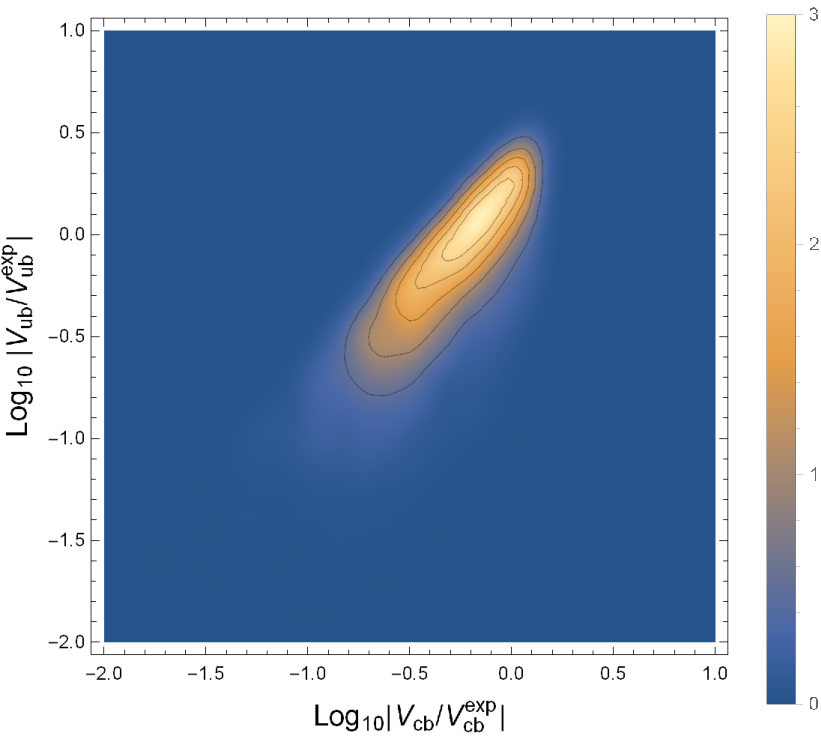}
\caption{Distribution of CKM matrix elements. \textbf{Left}: Plot of $|V_{us}/V_{us}^{exp}|$ versus $|V_{cb}/V_{cb}^{exp}|$. \textbf{Right}: Plot of $|V_{cb}/V_{cb}^{exp}|$ versus $|V_{ub}/V_{ub}^{exp}|$. The legend indicates the density height of the resulting 2-dimensional outcomes.}
\label{fig:CKM_sta}
\end{figure}

\begin{figure}[htb]
\centering
\includegraphics[width=0.4\textwidth]{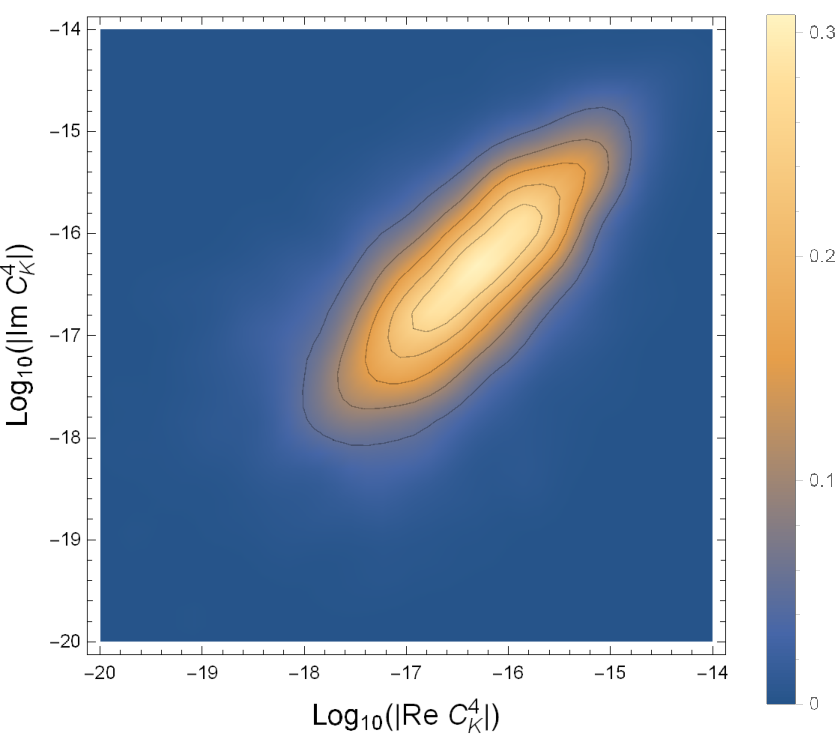} \hspace{0.5cm}
\includegraphics[width=0.4\textwidth]{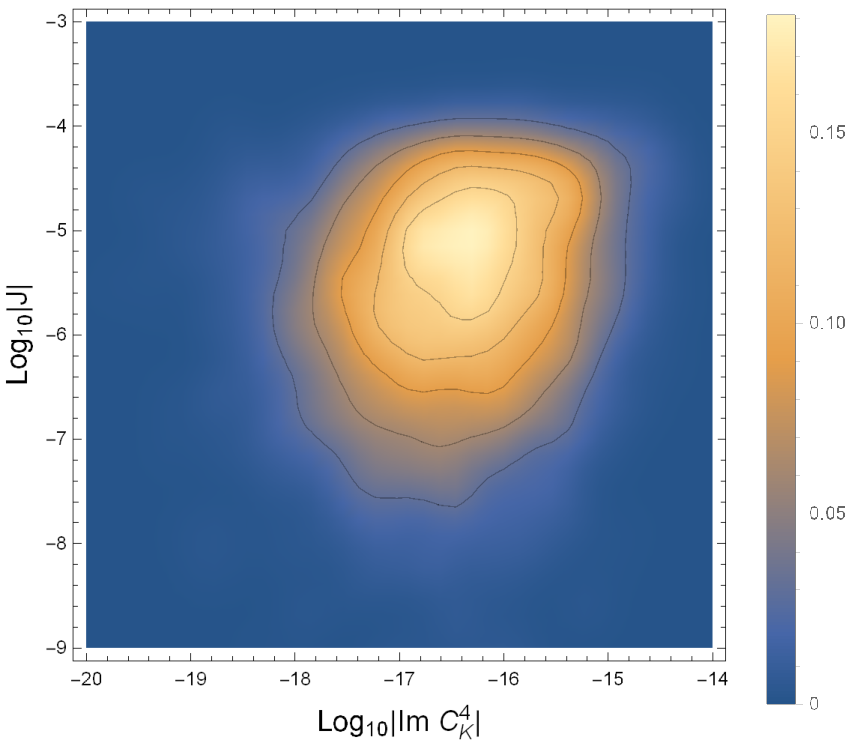}
\caption{CP violation quantities in this model. \textbf{Left}: Real part ($\text{Re}\, C_K^4$) versus imaginary part ($\text{Im}\, C_K^4$), in units of $\text{GeV}^{-2}$. \textbf{Right}: Imaginary part ($\text{Im}\, C_K^4$) versus the magnitude of the Jarlskog invariant ($|J|$). For illustration, we take $\mu_K = 2\,\mgev$ and $v_1 = 2\, \mpev$. The legend indicates the density height of the corresponding data points.}
\label{fig:FCNC_J}
\end{figure}

 Approximately $2\%$ of the randomly generated configurations exhibit a mild hierarchical structure, characterized by $0.3>|V_{us}| > 3|V_{cb}| > 15 |V_{ub}|$, $m_b > 5 m_s > 25 m_d$, and $m_t > 5 m_c > 25 m_u$. Statistics of the resulting CKM elements, $C_K^4$, and $J$ from 5000 such configurations are displayed in Fig.~\ref{fig:CKM_sta} and Fig.~\ref{fig:FCNC_J}.

From Fig.~\ref{fig:CKM_sta}, it is evident that the observed CKM mixing pattern falls within the expected range of this model, requiring minimal fine-tuning. Additionally, the absolute value of \(J\) is approximately correct, as illustrated in the right panel of Fig.~\ref{fig:FCNC_J}.

Regarding the low-energy \(\Delta S=2\) FCNC constraint, this model provides ample parameter space to accommodate realistic configurations, as illustrated in Fig.~\ref{fig:FCNC_J}. Notably, selecting \(v_1 = 10\) PeV instead of \(2\) PeV shifts the central value of \(|\text{Im}(C^4_K)|\) from approximately \(10^{-16.4} \text{GeV}^{-2}\) to \(10^{-17.8} \text{GeV}^{-2}\) (see Fig.~\ref{fig:K4_D4}), thereby easily circumventing the constraint \(\text{Im}(C^4_K) \in [-1.8, 0.9] \times 10^{-17} \text{GeV}^{-2}\) \cite{UTfit:2007eik}.

\section{Potential Experimental Signatures }
\label{sec:pheno}

\subsection{Phenomenology of the new gauge boson}
\label{sec:ZX_pheno}
Due to the gauge nature of the model, the existence of a new gauge boson, \(Z_F\), is inevitable, and its discovery would serve as compelling evidence. The mass of \(Z_F\) is an unknown parameter. If the theory remains perturbative, one can only infer that \(M_F \lesssim v_1\).
Note, however, that the low-energy FCNC four-fermion effective operators, mediated by tree-level flavor-crossing \(Z_F\) couplings, are determined solely by the \(U(1)_F\) charges and \(v_1\), and are independent of \(M_F\).

\begin{figure}[htb]
\centering
\includegraphics[width=0.4\textwidth]{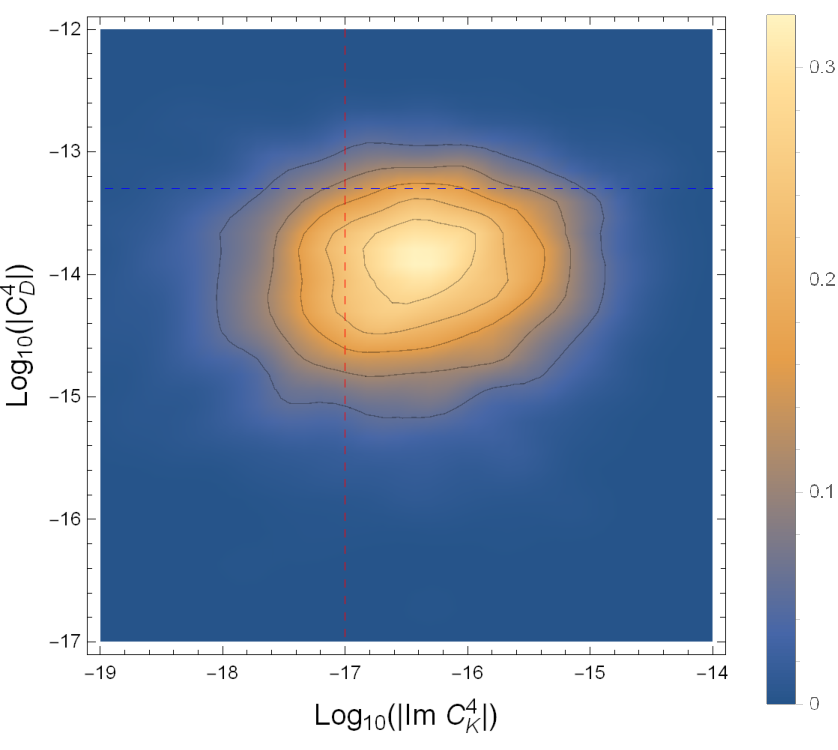} \hspace{0.5cm}
\includegraphics[width=0.4\textwidth]{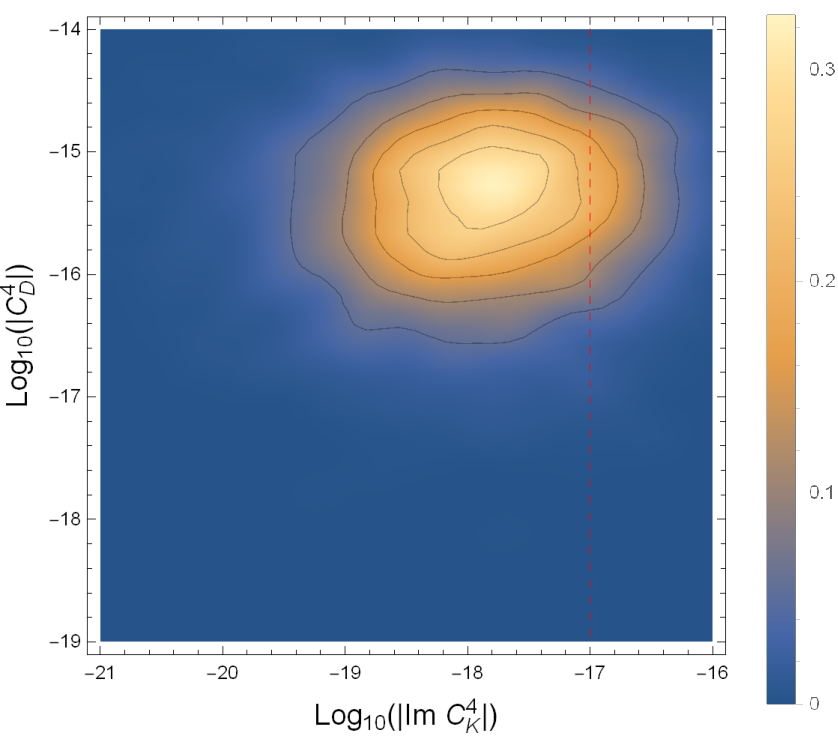}
\caption{Imaginary part of \( \Im C_K^4 \) versus absolute value of \( |C_D^4| \), in units of \(\text{GeV}^{-2}\).
The region to the right of the red dashed line represents the excluded region based on the current bound of \( \Im C_K^4 \).
The region above the blue dashed line is excluded by the current limit on \( |C_D^4| \).
\textbf{Left}: \(v_1 = 2 \) PeV. \textbf{Right}: Same plot as the left but with \( v_1 = 10 \) PeV.
The legend indicates the density height of the resulting 2-dimensional predictions.}

\label{fig:K4_D4}
\end{figure}

The value of \(v_1\) remains unknown. If \(v_1\) is relatively low, we might have a chance to detect the footprint of \(Z_F\) with improved low-energy FCNC experimental data. For instance, if \(v_1\) is \(2\,\text{PeV}\), as shown in the left panel of Fig.~\ref{fig:K4_D4}, the `natural' region for \(|C_D^4|\) is \(\gtrsim 10^{-15} \mgev^{-2}\) and for \(|\text{Im} \, C_K^4|\) is \(\gtrsim 10^{-18}\mgev^{-2}\), thus an order of magnitude improvement in experimental sensitivity could be sufficient. However, if \(v_1\) is pushed to \(10\,\text{PeV}\), as shown in the right panel of Fig.~\ref{fig:K4_D4}, the lower bound of the `natural' region for \(|C_D^4|\) shifts to \(\gtrsim 10^{-16}\mgev^{-2}\) and for \(|\text{Im} \, C_K^4|\) to \(\gtrsim 10^{-19}\mgev^{-2}\).

Next, we will discuss the phenomenologically interesting scenario where \(Z_F\) is relatively light, specifically \(M_F \lesssim 10\,\text{TeV}\), such that \(M_{1,3,\pm,N} \gg M_F\). In addition to contributing to FCNC \(\Delta F = 2\) transitions, \(Z_F\) might also exhibit interesting signatures at colliders in the near future.

If the SM fermion masses can be neglected, one can transition to the interaction basis to calculate the total decay width, which is proportional to \(\sum_f \left[ (Q_L^f)^2 + (Q_R^f)^2 \right]\). Using the \(U(1)_F\) charges from Table~\ref{table:SMparticle}, the decay width of \(Z_F\) can be expressed as:
\beq
\Gamma_F = \frac{151}{12\pi} M_F g_F^2 = 4.01 \frac{M_F^3}{v_1^2} = 8.02 \left( \frac{M_F}{20\,\text{TeV}} \right)^3 \left( \frac{2\,\text{PeV}}{v_1} \right)^2 \,\text{GeV}.
\eeq
Comparing this to its mass,
\beq
\frac{\Gamma_F}{M_F} = 4.01 \frac{M_F^2}{v_1^2}= 4.01\times 10^{-6}\left( \frac{M_F}{2\,\text{TeV}} \right)^2 \left( \frac{2\,\text{PeV}}{v_1} \right)^2 \,,
\eeq
indicating an extremely narrow resonance when the gauge boson is relatively light and directly produced at colliders.

It is essential to measure the coupling strengths of the SM fermions to distinguish this model from other \(Z'\) models. Notably, by construction, there are almost no flavor-diagonal couplings to \(\tau\), \(\nu_\tau\), top quark, and bottom quark. For the first and second generations, the actual flavor-dependent couplings depend on the specifics of the quark mass matrix diagonalization, and hence, there is no definitive prediction. However, we anticipate that the gross pattern should not vary significantly. For illustration, using the benchmark configuration discussed earlier, this model predicts the following ratio for the production cross-sections near the resonance of \(Z_F\) in different channels:
\beqa
\sigma(e^+e^- \to Z_F^* \to e\bar{e}) : \sigma(e^+e^- \to Z_F^* \to \mu\bar{\mu}) :
\sigma(e^+e^- \to Z_F^* \to s\bar{s}) : \sigma(e^+e^- \to Z_F^* \to d\bar{d}) \nonumber \\
: \sigma(e^+e^- \to Z_F^* \to u\bar{u}) : \sigma(e^+e^- \to Z_F^* \to c\bar{c}) \simeq 4.8 : 4.8 : 3.1 : 3.1 : 1 : 1
\eeqa
with
\beq
\sigma(e^+e^- \to Z_F^* \to \mu\bar{\mu})\simeq \left( \frac{61}{151}\right)^2\frac{6\pi}{M_F^2} =1.19 \times \left( \frac{\mtev}{M_F}\right)^2\,\mbox{nb}
\eeq

\begin{figure}[htb]
\centering
\includegraphics[width=0.45\textwidth]{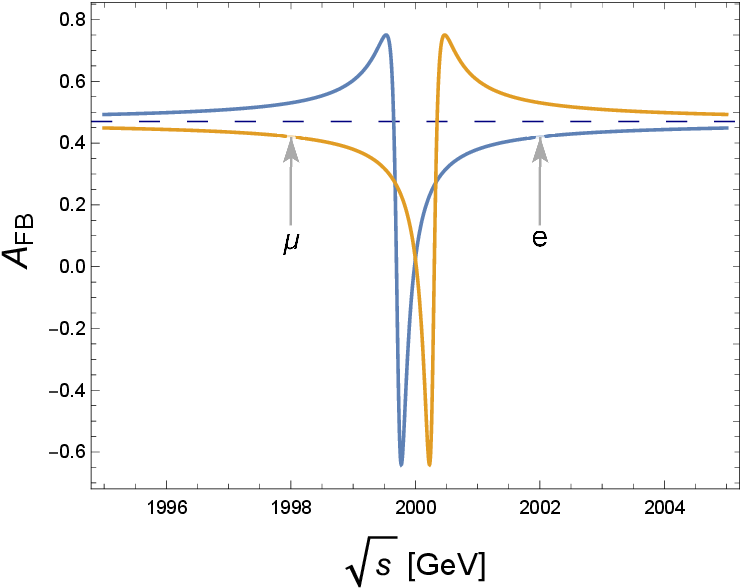}
\caption{Forward-backward asymmetry near the \(M_F = 2\,\text{TeV}\) resonance with \(v_1\) set to \(2\,\text{PeV}\). The dashed line represents \(A_{FB}^\tau\), the SM prediction, as the \(\tau\) lepton does not couple to \(Z_F\).}
\label{fig:AFB}
\end{figure}

Moreover, the forward-backward asymmetries of electrons and muons are predicted to deviate significantly from the SM predictions near the \(Z_F\) resonance, whereas the \(\tau\) lepton does not exhibit such modifications, as shown in Fig.~\ref{fig:AFB}. Since the first- and second-generation charged leptons possess opposite \(U(1)_F\) charges, their asymmetry lineshapes form a mirror image with respect to the resonance. Once observed, this phenomenon would provide a convincing indication of the existence of this specific gauged \(U(1)\) symmetry.

Before closing this subsection, we remark that although only the SM Higgs and \( S_1 \) develop nonzero VEVs, the SM Higgs can still mix with \( H_1 \) through the cubic coupling \(\mu_{0,1,s} \overline{H_1} H S_1\). Using the benchmark point, the mixing angle is determined by
\beq
\theta_H = \frac{1}{2} \tan^{-1} \left( \frac{\mu_{0,1,s} v_1}{M_1^2 - M_h^2} \right) = 0.006\,,
\eeq
where \( M_h \) is the mass of the SM Higgs. This results in a universal suppression factor, \(\cos \theta_H\), applied to all SM Higgs couplings, leading to an \({\cal O}(10^{-5})\) deviation from the SM prediction. Such a small deviation poses significant experimental challenges for detection.

\subsection{$\tri a_e$, $d_e$, and Effective Electron Yukawa}
\label{sec:delta_m_E}
Here, we provide a more detailed examination of the electron mass generation.

The cubic coupling $\bar{H_3} H_- S_1^*$
results in three physical charged scalars participating in the one-loop diagram ( Fig.~\ref{fig:e_mass} ) responsible for electron mass generation. We label these charged scalars as $\eta_i^+$
for $i=1,2,3$. These charged scalars are related to $\tilde{\eta}=\{ H_3^+,H_-^+, C_4^+\}$
 via an orthogonal transformation, such that
$ \tilde{\eta}_a = G_{ai} \eta_i^+$. In this context, \(G\) represents the diagonalization matrix for the charged scalar mass-squared matrix

\beq
\begin{pmatrix}
    M_3^2 & \mu_{-,3,s}\frac{v_1}{\sqrt{2}} & -\frac{\lambda_e v_1 v_0}{2} \\
    \mu_{-,3,s}\frac{v_1}{\sqrt{2}} & M_-^2 & 0 \\
    -\frac{\lambda_e v_1 v_0}{2} & 0 & M_4^2
\end{pmatrix}
\eeq
in the basis of \(\tilde{\eta}\).

 It is observed that $H_3^+$ and $H_-^+$
 exhibit sizable mixing, whereas the mixing between  $H_3^+$ and $C_4^+$ is much smaller.
And the interaction between electron and the vector fermion $N$ can be expressed
as ${\cal L} \supset \bar{N}(\alpha_i \PL +\beta_i \PR) e\, \eta_i^+ + H.C.$, where
$\alpha_i= y^e_3 G_{1i}$ and $\beta_i= y^e_C G_{3i}$.
The one-loop contribution to the electron mass is finite and is expressed as
\beq
\delta m_e = {M_N\over 16\pi^2 }\sum_{i=1}^3 \alpha_i \beta_i^* {\rho_i \ln \rho_i \over 1-\rho_i}\,,
\eeq
where  $\rho_i= (M_{\eta_i}/M_N)^2$.

Furthermore, the one-loop BSM contribution to the anomalous magnetic moment can be calculated as follows \cite{Chang:2021axw}:
\beq
\Delta a^{BSM}_e = \frac{1}{16 \pi^2}\frac{m_e}{M_N}\sum_{i=1}^3 \mathcal{R}(\alpha_i \beta_i^*) I_2(\rho_i)\,,
\eeq
where
\beq
I_2(x) = \frac{1-x^2+2x \ln x}{(x-1)^3}\,.
\eeq

Considering the radiative generation of the electron mass, \(m_e = \delta m_e \simeq 0.511 \text{ MeV}\), which is a physical and real number, the phase of \(y^e_3 (y^e_C)^*\) must be removed through field redefinition. As a robust prediction, this model asserts \(\Delta a^{BSM}_e < 0\), although the one-loop correction, \(|\Delta a^{BSM}_e| \simeq \frac{m_e^2}{\Lambda^2} \sim \mathcal{O}(10^{-21})\), is too small to explain the observed value\cite{Parker:2018vye}. Additionally, the one-loop electric dipole moment of the electron, \(\propto \Im[y^e_3 (y^e_C)^*] = 0\), vanishes.

In this model, the new BSM fields to which the electron couples are \(N\), \(C_4\), \(H_3\), and \(Z_F\). Moreover, there are only four distinguished topologies of two-loop diagrams~\cite{Chang:2000wf,Chang:2004pba}. An exhaustive but straightforward examination of all two-loop Feynman diagrams confirms \(d_e = 0\) up to the two-loop level, which automatically mitigates the CP problem, rendering the BSM contribution to \(d_e\) negligible.

One can proceed to calculate the effective Yukawa coupling of the electron to the SM Higgs, $h^0_{SM}$. This calculation incorporates additional parameters in the scalar sector and may lead to a detectable deviation from the SM prediction, $y_e^{SM}= m_e/v_0$.
Here, we denote the cubic coupling as $ \mu_{ij} \eta_i^* \eta_j h^0_{SM}$, involving a dimensionful parameter $\mu_{ij}$.

The effective electron Yukawa coupling can then be determined as
\beq
y_e^{eff}=\sum_{i j} {\alpha_i^* \beta_j\over 16 \pi^2} \frac{\mu_{ij}}{M_N}  I_3(\rho_i,\rho_j, \rho_q)
\eeq
where $\rho_q= q^2/M_N^2$, with $q$ being the 4-momentum of the SM Higgs, and the electron mass is ignored.  The integral function is defined as
\beq
I_3(a,b,c)= \int^1_0 dx \int^{1-x}_0\! dy\, {1 \over 1-x-y+ x a+y b-x y c}\,.
\eeq
Assuming that $q^2 \ll M_N^2$,  the integral simplifies to
\beq
I_3(a,b,0) = {1\over a-b}\left( {a \ln a \over a-1} - {b \ln b \over b-1} \right)\,,
\eeq
and it approaches the limit  ${ a-1-\ln a \over (a-1)^2}$ when $a=b$.

Regarding the cubic couplings $\mu_{ij}$, the strengths of the quartic couplings $|H_3|^2 |H|^2$, $|H_-|^2 |H|^2$, and $|C_4|^2 |H|^2$ also contribute. Specifically, the cubic couplings are given by
\begin{equation}
\mu_{ij} = \frac{\lambda_e v_1}{2} G_{1i} G_{3j} + \lambda_3 v_0 G_{1i} G_{1j} + \lambda_- v_0 G_{2i} G_{2j} + \lambda_4 v_0 G_{3i} G_{3j}\,,
\end{equation}
where $\lambda_3$, $\lambda_-$, and $\lambda_4$ are the corresponding quartic coupling constants. However, their significance is highly suppressed by the factor $v_0/v_1$.

Unlike the cases studied in \cite{Fraser:2014ija, Chang:2022eft, Chang:2022pue}, there are no tree-level contributions to the electron mass, and the mixing between \( H_3^+ \) and \( C_4^+ \), \( \sim (v_1 v_0 / \Lambda^2) \simeq 10^{-6} \), is exceedingly small. Consequently, the resulting effective electron Yukawa coupling is expected to match the SM value. To verify this, we conducted a numerical scan to explore the effective electron Yukawa coupling to the SM Higgs. By requiring the correct electron mass, we allowed the parameter \( M_N \) to vary between \( 1\, \mpev \) and \( 50\, \mpev \), while \( \lambda_3 \), \( \lambda_- \), and \( \lambda_4 \) were varied between \( 0.01 \) and \( \sqrt{4\pi} \). As expected, the deviation from the SM prediction is undetectable.


\section{Conclusion}
\label{sec:summary}

In summary, our exploration of an extension of the SM provides a novel perspective on the origins of charged fermion masses. This extension, characterized by an anomaly-free gauge \( U(1)_F \) horizontal symmetry, introduces additional scalar doublets, scalar singlets, and vector-like fermionic singlets, while avoiding exotic chiral fermions. This setup effectively addresses the observed hierarchies and mixing patterns among SM charged fermions.

Our proposed model features a flavor non-universal, anomaly-free \( U(1)_F \) charge assignment that incorporates both radiative generation and FN mechanisms. This charge assignment ensures that third-generation quarks and lepton acquire masses at leading order. The expanded scalar sector, comprising un-Higgsed doublets and Higgsed singlet, is instrumental in implementing the FN mechanism for second-generation masses and facilitating the spontaneous symmetry breaking of the gauge \( U(1)_F \) symmetry.

In conjunction with the top quark, this extended scalar sector enables the generation of first-generation quark masses through one- and two-loop quantum corrections, offering a compelling resolution to the flavor puzzle. Additionally, the introduction of vector-like singlet fermions and a charged scalar singlet is essential for the one-loop generation of the electron mass, while ensuring a vanishing electric dipole moment at the two-loop level and preventing charged lepton flavor violation to all orders.

Numerical studies validate the framework’s ability to naturally reproduce a realistic charged fermion mass spectrum and the observed CKM mixings with minimal fine-tuning. The five orders of magnitude differences observed in the SM Yukawas can be substantially reduced to two in this model. Potential experimental signatures of this model include a narrow resonance if the new gauge boson is light (\(\lesssim 10\) TeV), with measurable predictions for \(\text{Br}(Z_F \rightarrow f \bar{f})\) and specific patterns of \(A_{FB}^e\) and \(A_{FB}^\mu\) at future colliders, distinguishing this model from other \(Z'\) models. Additionally, if the VEV of \(U(1)_F\) is low (approximately 2 PeV), our numerical scan suggests that FCNC couplings of \(Z_F\) might be detectable in \(K^0\overline{K^0}\) and \(D^0\overline{D^0}\) mixings, specifically \( |\Im C_K^4| \gtrsim 10^{-18} \ \text{GeV}^{-2} \) and \( |C^4_D| \gtrsim 10^{-15} \ \text{GeV}^{-2} \).

In conclusion, this gauge horizontal model not only addresses existing flavor puzzles within the SM charged fermions but also provides an economical UV-complete implementation of the FN mechanism at relatively low energies (\( v_1 \sim \mathcal{O}(\text{PeV}) \) and \( \Lambda \sim \mathcal{O}(10 \text{PeV}) \) are feasible). Future studies will work on integrating the neutrino sector into our model, resulting in a comprehensive framework for all fermion masses and mixings. We also propose enhanced experimental investigations to test our predictions, which could yield valuable insights into flavor physics and affirm or question the validity of our proposed model.

\section*{Acknowledgments}
This research is supported by NSTC grant 112-2112-M-007-012, Taiwan

\bibliography{gaugedWF_Ref}
\end{document}